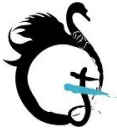

Article

# Singularity and Coordination Problems: Pandemic Lessons from 2020

**Nicholas Kluge Corrêa[1,\*], Nythamar de Oliveira[1]**

[1]*Graduate Program in Philosophy of the Pontifical Catholic University of Rio Grande do Sul (Brazil) – Av. Ipiranga, 6681 - Partenon, Porto Alegre - RS, 90619-900.*

## Abstract

*Are there any indications that a Technological Singularity may be on the horizon? In trying to answer these questions, the authors made a small introduction to the area of safety research in artificial intelligence. The authors review some of the current paradigms in the development of autonomous intelligent systems, searching for evidence that may indicate the coming of a possible Technological Singularity. Finally, the authors present a reflection using the COVID-19 pandemic, something that showed that global society's biggest problem in managing existential risks is its lack of coordination skills as a global society.*

## Keywords

Singularity, Artificial Intelligence, Existential Risk, Coronavirus Pandemic

## Singulitarianism and Safety

Artificial Intelligence (AI) research is an interdisciplinary endeavor by nature, given the various fields that participate and benefit from its development. When we talk about AI, either in the context of computer science (Sutton & Barto,1998; Russell & Norvig, 2003; Wang, 2019) or in the study of the philosophy of the mind (Searle, 1980; Haugeland, 1985; Newell, 1990; Chalmers, 2010), a certain dichotomy is utilized to classify two different types of AI: Narrow (1) and General (2) (Shane & Hutter, 2007):

1.  *Narrow intelligence*: Also known as "weak" AI, narrow AI is how we define intelligent systems that we are used to interacting within our daily lives. Such systems are only proficient in specific tasks and unable to generalize their skills to domains outside their training environment;

2.  *General intelligence*: also referred to as "strong" AI, or artificial general intelligence (AGI), which consists of an hypothetical intelligent system capable of solving many types of problems proficiently, in any domain, or at least in a wide range of domains.

AGI (depending on how we interpret the word "general") would be something capable of covering all possible tasks. Those that humans are specifically good at. Those that animals are specifically good at. And all that goes beyond our imagination (Chollet, 2019). However, a more modest definition would be to define an AGI as a system that exceeds human cognitive ability in any domain of interest (Müller & Bostrom, 2016).

Moravec (1998, p. 10) proposes an analogy where the advancement of AIs capabilities are compared to a "flood." Fifty years ago, tasks previously performed exclusively by humans (e.g., human calculators) were "flooded" and replaced by the use of autonomous systems. We are increasingly taking refuge in the high peaks of the cognitive landscape, still reserved exclusively for us, while lower regions continue to be flooded.

The authors' objective in this essay is to explore the following idea, "What if we are successful in developing AGI"? Vinge (1993) uses the term "Singularity" to define artificial intelligent systems/agents that have surpassed human intelligence. At the same time, "Singulitarianism" is the name used to describe the Transhumanist strand where the possibilities and consequences of creating a Technological Singularity in the medium-long future are discussed. Given this possibility, an active response is necessary to ensure that such an event is beneficial to our

---
* *Corresponding author*.
  *E-mail addresses*: nicholas.correa@acad.pucrs.br (N. K. Corrêa), nythamar@yahoo.com (N. de Oliveira).

           61

society (Kurzweil, 2005; Naude, 2009; Chalmers, 2010; Lombardo, 2012; Tegmark, 2017).

Irving J. Good was one of the first academics to speculate on the possibility of an "ultraintelligent machine" (Singularity):

> Let an ultraintelligent machine be defined as a machine that can far surpass all the intellectual activities of any man however clever. Since the design of machines is one of these intellectual activities, an ultraintelligent machine could design even better machines; there would then unquestionably be an "intelligence explosion", and the intelligence of man would be left far behind […] Thus the first ultraintelligent machine is the last invention that man need ever make, provided that the machine is docile enough to tell us how to keep it under control. It is curious that this point is made so seldom outside of science fiction. It is sometimes worthwhile to take science fiction seriously. (Good, 1965, p. 33)

However, would there be any indication that an intelligence explosion is something, however unlikely, still possible? Perhaps.

One can already find in the literature the first indications of autonomous systems assisting in the development of other autonomous systems. Zoph and Le (2017) proposed an autonomous technique for the development of artificial neural network architecture. According to the authors: "our method, starting from scratch, can design a new network architecture that rivals the best architecture invented by man" (Zoph & Le, 2017, p. 1). The authors developed their model using Reinforcement Learning (RL) to train their "architect" system of artificial neural networks. RL is one of the paradigms in the area of machine learning, where artificial agents must act in the environment that they are embedded and find an action policy that maximizes the cumulative reward return for a given reward function (i.e., the goal) (Russell & Norvig, 2003).

Reward functions are a mathematical representation of the preferences that guide the behavior of agents operating by RL, where, for example, a cleaning robot can maximize a function that assigns high reward to world-states with "little dirt on the floor" and a low reward to world-states where the floor is dirty.

Many of the models used to study idealized rational agents (e.g., Expected Utility Theory) provide convincing arguments that any rational agent with consistent preferences should act as an expected utility maximizer (Von Neumann & Morgenstern, 1944). However, within the framework of expected utility theory, there are corollary results that seem to refer to the concern of Good, quoted above: "as long as the machine is docile enough to tell us how to keep it under control" (Good, 1965, p. 33).

Stephen Omohundro (2008) cites some characteristics that we should expect these types of agents (expected utility maximizers/RL agents) to possess, and Bostrom (2014, chapter 7, p. 110-112) popularized Omohundro's arguments in two theses:

- *Instrumental Convergence Thesis*: Intelligent agents can have a wide range of possible terminal goals. However, certain instrumental goals can be pursued by almost all intelligent agents. Since these goals are means for the achievement of almost any terminal goal;

- *Orthogonality Thesis*: Analogous to Hume's Guillotine (Is-Ought Gap), the orthogonality thesis dictates that ethical pronouncements for what should be cannot be achieved through factual analysis. Thus, both concepts (reason and morality) would be independent. As a result, terminal goals and levels of intelligence can freely vary in orthogonal axis.

Turner, Smith, Shah and Tadepalli (2020) generalized the conjectures made by Omohundro and Bostrom in what the authors call the "Power-Seeking Theorems." In them, it is demonstrated that within the formalism of Markov decision processes (MDP), most reward functions encourage "power-seeking behavior." Power, in MDPs, is the ability to achieve goals in general. It is instrumentally convergent to a wide range of reward functions to seek power. A corollary of the results demonstrated by Turner et al. (2020) is that even in simplified conditions, most reward functions induce power-seeking behavior, something that may cause safety problems involving the interaction of humans and AI (e.g., an AGI how has incentives to avoid its shutdown).

In light of all these arguments, which date back to the early days of AI research, security issues have increasingly been cited in the literature. AI ethics, a sub-area of applied ethics concerned with adding moral behavior to machines and regulating the use of artificial intelligence, has gained a significant increase in popularity in the last two decades (Jobin, Ienca, & Vayena, 2019; Jurić, Šandić, & Brcic, 2020). At the same time, important philosophical and technical questions are raised in the context of AI safety, e.g., Corrigibility: how to correct/terminate potentially faulty agents that have a strong instrumental incentive to preserve their terminal goals (Soares, Fallenstein, Yudkowsky, & Armstrong, 2015; Amodei et al., 2016)?

One can also find in the literature several research agendas, where different types of ethical, technical, and social problems are discussed (Russell, Dewey, & Tegmark, 2015; Taylor, Yudkowsky, Lavictoire, & Critch, 2016; Tegmark, 2016; Soares, 2016; O'Keefe et al., 2020; ÓhÉigeartaigh, Whittlestone, Liu, Zeng, & Liu 2020; Hagendorff, 2020; Corrêa & De Oliveira, 2021). For example, how will we remedy the negative economic impacts of AI, such as mass automation and unemployment (Frey & Osborne, 2013)? How can we prevent the automation of jobs to increase the inequality relation among classes, genders, and races (Brynjolfsson & McAfee, 2014)? Should autonomous weapons be banned (Docherty, 2012)?

At one end of the spectrum, we find research involving existential risks, i.e., the study of possible threats at the extinction level imposed by present or future technology. Research centers such as the Centre for the Study of Existential Risk (Cambridge), Future of Life Institute (Harvard/MIT), Future of Humanity Institute (Oxford), Machine Intelligence Research Institute, and the Center for Human-Compatible AI (both at Berkeley) seek to develop strategies to mitigate these possible threats.

Throughout this article, more arguments will be proposed to justify the type of research the authors address (AI Ethics and AI Safety). In the next section, two types of "scenarios" of how society could come to "lose control" are presented.

## AI Takeoff

During the 20th century, a technological race led to the mass production of systems that we did not yet have a complete understanding of. Something that caused various side effects, such as accidents (Chernobyl disaster), the creation of weapons of mass destruction (Cold War), and even the use of these weapons against human society itself (bombing of Hiroshima and Nagasaki).

Certainly that there are various pressures to develop high-performance AI, given its ability to provide the organization controlling it with a considerable strategic advantage. The great instrumental value in being the first global actor to control AGI may cause the same kind of technology race that we experienced in the mid-20th century. That is: "while X invests in the development of AI, Y will do as well."

Another reason to be cautious in our technological advances in the area of AI is that for an intelligent autonomous system to pose a potential danger to our society, it doesn't need to be smarter than us. It just needs to be more capable regarding certain types of tasks. Barrett and Baum (2017) explore two main reasons that would cause an artificial intelligence to represent a considerable danger to our society, reasons of *capability* (1) and *value* (2).

1. Intelligent artificial agents can pose a danger to human well-being because of their extremely refined ability, or some aptitude, with which we cannot compete;

2. Intelligent artificial agents can develop goals and objectives that diverge from us humans, and in pursuing them, cause damage to our society.

ASI-PATH (Artificial Super Intelligence Pathway) is a model of how an AGI, becoming super intelligent through recursive self-improvement, could come to cause a catastrophe (Barrett & Baum, 2017). This model suggests scenarios where an AI, after achieving some DSA (Decisive Strategic Advantage), e.g., advances in nanotechnology, biological engineering, or robotics, could come to achieve a considerable level of control over the environment.

What would be a "good example" of a DSA?

Given our reliance on autonomous systems integrated with the Internet, one potentially damaging DSA would be to conduct cyberattacks on vital structures of our infrastructure, such as electricity distribution and telecommunications networks. In 2017, the crypto-ransomware "WannaCry" broke into systems in more than 99 countries, even affecting the public health system of certain governments. More than 75,000 ransom demands were made, making it one of the most damaging cyberattacks in history (Larson, 2017). This would be a possible DSA of an AI, i.e., the ability to execute cyberattacks there our infrastructure in a way that we cannot remediate in time.

ASI-PATH provides an intuitive diagram where several events (i.e., security breaches) must occur for a catastrophe involving AGI to occur. Initially, an AI, also called a seed AI, must first become an AI with some DSA, and at the same time, certain security measures must fail:

- Failure of AI confinement;
- Unsuccessful value alignment;
- AI's goals diverge from ours.

To those interested, Sotala (2018, p. 317) provides a simplified overview of ASI-PATH in his paper "Disjunctive scenarios of catastrophic AI risk." According to Barrett and Baum (2017), the arguments raised by the instrumental convergence thesis and the orthogonality thesis (i.e., power-seeking and terminal goal divergence) are some of the reasons that could lead a Singularity to engage in hostile actions against humanity

The scenarios explored in the literature, where a seed AI is capable of becoming a Singularity, are usually characterized in two different types of takeoffs: *Fast* and *Slow*. Fast takeoffs suggest situations where a drastic takeover occurs, where abruptly we would be surprised by an entity much more capable, with possibly unknown objectives, inserted and sharing the same environment as us. In contrast, we have slow takeoffs, which are a much more realistic possibility.  It would occur gradually as the human species becomes more and more dependent, and in a way, under the control of advanced AI systems (Sotala, 2018). Such questions raise concerns, especially in the area of ethics and morals. Old questions are now reexamined in a new light, and even with a new sense of urgency. For AI development to be done in a way that minimizes the risk of existential threats to humanity, some questions still unanswered are:

1. What strategies and policies should we adopt to ensure that the goals of advanced artificial agents are aligned with our interests?

2. What restrictions to this project should we impose to ensure a beneficial outcome?

3. Would there be predictions of when an AGI could be achieved?

In the following sections, the authors answer some of these questions, starting with the last one.

**AGI On the Horizon?**

Technological forecasting is highly complex, and how pessimistic or optimistic we should be is not clear. Several scientists proposed predictions that ended up being wrong. The nuclear physicist Ernest Rutherford, in 1933, said that anyone who defended the possibility of one day extracting the energy contained in the atomic nucleus was "talking moonshine." Also, in 1896, Lord Kelvin said to not have the "slightest molecule of faith" in any type of air navigation besides ballooning. Could skeptics about the emergence of an AGI be victims of the same fate?

Experts in the development of artificial intelligence predict that within 10 years many human activities will be surpassed by machines in terms of efficiency (Grace, Salvatier, Dafoe, Zhang, & Evans, 2017). A survey was conducted with several experts (N = 170) by Müller and Bostrom (2016) to assess the progress in AI research and prospects for the future. The survey showed that on average, there is a 50% chance that high-level machine intelligence will be achieved between 2040 and 2050, with a 90% probability by 2075. According to the interviewees' opinion, AI will outperform human performance between 2 (10% chance) and 30 years (75% chance) (Müller & Bostrom, 2016). In Müller and Bostrom's (2016) survey, 33% of respondents classified this development in AI as "bad" or "extremely bad" for humanity.

In a similar survey conducted by Grace et al. (2017), the researchers interviewed (352 participants of the 2015 NIPS and ICML conferences) believe that AI will outperform human performance in all tasks in 45 years, with a 50% chance, and automate all human work in up to 120 years. In the research of Grace et al. (2017), when those evaluated were asked the question "Does Stuart Russell's argument for why highly advanced AI might pose a risk point at an important problem?", 70% of respondents answered, "Yes" (Grace et al., 2017, p. 13).

Besides the opinion of specialists in the field, another type of evidence that we can use to infer the possibility of a technological Singularity is how the economic growth rate has behaved during the history of human civilization, and how it's related to technological improvement.

One of the most popular models found in the literature on our economic growth, from the Neolithic Revolution to the 21st century, is the growth model proposed by Michael Kremer (1993). Kremer's model is based on the following simple argument: Two heads think better than one, i.e., economic growth is driven by people having new ideas, and the more people, the greater the chance of new ideas.

For Kremer (1993), the total annual economic output is a function of the size of the population and the level of technology of this population. Kremer also assumes that if there are no changes in technology (e.g., advances in agriculture), if we have double the number of people working in a given piece of land, this will not necessarily double the food produced on this land. Thus, population growth depends on technological progress. At the same time, technological growth depends on population size, which makes the rate of population growth, technological progress, and economic production factors dynamically dependent on each other.

One property of Kremer's growth model is that it indicates a form of hyperbolic growth, and hyperbolic curves tend to infinite values, i.e., at some point, we will reach some form of singularity. This model also suggests that such forms of growth should be separated when we reach a maximum population growth rate of 2100, with a global population between 9.6 billion and 12.3 billion people (Gerland et al., 2014). When this occurs, technological progress will no longer impact the global population. However, this does not mean that technological progress will stagnate.

This type of model is sometimes referred to as the Hyperbolic Growth Hypothesis (HCH). HCH is one of the most accepted economic growth models by the macroeconomic community, and serves as the basis for other theories such as the Unified Growth Theory (Taagepera, 1979; Korotayev, Malkov, & Khaltourina, 2006; Oded, 2011; Jones, 2013). Other authors also suggest a disassociation between population growth and economic/technological progress (Yudkowsky, 2013; Bostrom, 2014; Nordhaus, 2015; Agrawal, Gans, & Goldfarb. 2017). Thus, when high levels of automation are achieved, economic growth rates will become radically higher, producing more and more technological progress.

Could this type of economic growth help the development of an AGI? Levin and Maas (2020) argue that when research involving advanced AI development is sufficiently theorized, efforts similar to the historic Manhattan Project could accelerate this project. At this point, international cooperation can change dramatically, causing implications for the stability of AI governance. At the time of the Apollo and Manhattan Projects, the U.S. government dedicated 0.4% of its GDP to accelerate the achievement of its objectives. This would currently amount to an annual budget of $80 billion to AGI R&D (Stine, 2009). A budget much larger than what was needed to accomplish some of the greatest technological achievements of the 21st century:

1. The Large Hadron Collider (HHC) at CERN (Conseil Européen pour la Recherche Nucléaire), took 10 years to build, at an annual cost of $475 million (Knapp, 2012);

2. The LIGO (Laser Interferometer Gravitational-Wave Observatory), had a total construction cost of US$ 33 million (Castelvecchi, 2015);

3. ITER (International Thermonuclear Experimental Reactor), one of the latest promises for clean and sustainable energy (a Tokamak nuclear fusion experimental reactor), is expected to be ready in 12 years at an annual cost of $2 billion (Fountain, 2017).

One can see that neither of the projects mentioned above has received as much economic investment as the one dedicated to the Apollo and Manhattan projects (0.4% of the U.S. government's annual Gross Domestic Product), something that also explains the impressive speed with which the goals of both projects were achieved. Even so, significantly less investment did not prevent major scientific discoveries, e.g., decoding of the human genome and the detection of gravitational waves. Thus, it seems feasible to state that: When we have a robust enough theoretical understanding of the computational and cognitive processes responsible for the development of AGI, a Singularity may very well be "a Manhattan Project" away.

Currently, there are several active projects to develop AGI. Baum (2017) identified 45 research and development projects intending to develop advanced artificial intelligence. The results of Baum's research are summarized in the table (Table 1) below:

**Table 1:** Advanced AI R&D projects.

| PROJECT | COUNTRY | INSTITUTION | MILITARY TIES | SAFETY ENGAGEMENT |
|---------|---------|-------------|---------------|-------------------|
| ACT-R | USA | Carnegie Mellon University | Yes | Not specified |
| AERA | CH | Reykjavik University | No | Active |
| AIDEUS | RUS | AIDEUS | Not specified | Active |
| AIXI | AUS | Australian National University | Not specified | Not specified |
| AIW | SE | Chalmers University of Technology | No | Not specified |
| Animats | SE | Chalmers University of Technology | No | Not specified |
| Baidu Research | CN | Baidu | Not specified | Not specified |
| Becca | USA | Becca | Not specified | Not specified |
| Blue Brain | CH | École Polytechnique Fédérale de Lausanne | Not specified | Not specified |
| CN Brain Project | CN | Chinese Academy of Sciences | Not specified | Not specified |
| CLARION | USA | Rensselaer Polytechnic Institute | Yes | Not specified |
| CogPrime | USA | OpenCog Foundation | Not specified | Active |
| CommAI | USA | Facebook | Not specified | Moderate |
| Cyc | USA | Cycorp | Yes | Not specified |
| DeepMind | UK | Google | Not specified | Active |
| DeSTIN | USA | University of Tennessee | Not specified | Not specified |
| DSO-CA | SG | DSO National Laboratories | Yes | Not specified |
| FLOWERS | FR | Inria and ENSTA ParisTech | Not specified | Active |
| GoodAI | CZ | GoodAI | Not specified | Active |
| HTM | USA | Numenta | Not specified | Non-existent |
| HBP | CH | École Polytechnique Fédérale de Lausanne | No | Not specified |
| Icarus | USA | Stanford University | Yes | Not specified |
| Leabra | USA | University of Colorado | Yes | Not specified |
| LIDA | USA | University of Memphis | Yes | Moderate |
| Maluuba | CA | Microsoft | Not specified | Not specified |
| MicroPsi | USA | Harvard University | Not specified | Not specified |
| MSR AI | USA | Microsoft | Not specified | Not specified |
| MLECOG | USA | Ohio University | Not specified | Not specified |
| NARS | USA | Temple University | Not specified | Active |
| Nigel | USA | Kimera | Not specified | Not specified |
| NNAISENSE | CH | NNAISENSE | Not specified | Not specified |
| OpenAI | USA | OpenAI | Not specified | Active |
| Real AI | CN | Real AI | Not specified | Active |
| RCBII | CN | Chinese Academy of Sciences | Not specified | Not specified |
| Sigma | USA | University of Southern California | Yes | Not specified |
| YesA | AT | Vienna University of Technology | Not specified | Not specified |
| SingularityNET | CN | SingularityNET Foundation | Not specified | Not specified |
| SNePS | USA | State University of New York | Yes | Not specified |
| Soar | USA | University of Michigan | Yes | Not specified |
| Susaro | UK | Susaro | Not specified | Active |
| TAIL | CN | Tencent | Not specified | Not specified |
| UAIL | USA | Uber | Not specified | Not specified |
| Vicarious | USA | Vicarious | Not specified | Moderate |
| Victor | USA | Cifer | Not specified | Non-existent |
| WBAI | JP | Whole Brain Architecture Initiative | Not specified | Active |

In Table 1, we can see that of the projects reviewed, ten have links with the military (nine working for the U.S. government, and one for the government of Singapore), while only four reportedly have no links to the military industry. All other projects do not specify their association with military agencies. Besides, of the 45 projects reviewed, only 13 have active/moderate involvement with the area of AI safety, while two of the projects reviewed (Hierarchical Temporal Memory and Victor) disregard the need for security measures entirely.

Jeffrey Hawkins, the leading researcher of the HTM (Hierarchical Temporal Memory) project, dismisses concerns related to advanced AI, stating, "I do not see machine intelligence representing any threat to humanity" (Hawkins, 2017, para. 3). And according to the 2AI Labs website, researchers give the following statement on risk scenarios involving advanced AI, "We think this is all crazy talk" (2AI Team, 2016, para. 2). The remaining 30 projects do not specify any type of research focused on the area of AI safety.

For those who follow the recent advances in the field of AI, it is known that one of the major paradigms of research today involves the problem of natural language processing (NLP), and the use of a new form of architecture called "Transformer", proposed by Vaswani et al. (2017) in their seminal work, *"Attention is all you need"*. Currently, systems based on the transformer architecture are the new paradigm in NLP tasks, reaching the highest records in the GLUE benchmark (General Language Understanding Evaluation) in tasks such as translation, text summarization, text classification, and Q&A.

Transformer models such as ELMo (Peters et al., 2018), BERT (Devlin, Chang, Lee, & Toutanova, 2019), GPT (Radford et al., 2019, Brown et al., 2020), and T-NLG (Corby, 2020), represent some of the most successful models in tasks involving NLP. GPT-2 has been used to generate several recent patent claims (Lee & Hsiang, 2019), while the GPT-3 language model is capable of generating samples of texts (e.g., poems, articles, news) from which human evaluators have difficulty distinguishing from human-produced text.

Besides achieving state-of-the-art in several tasks involving natural language processing, GPT-3 may be one of the first AI models to show signs of generalization. For example, even though it has never been trained to perform mathematical operations (GPT-3 is "only" a language model), GPT-3 is capable of answering complex questions involving 3-digit arithmetic. GPT-3 is also able to demonstrate sophisticated physical/scientific reasoning when evaluated in the PIQA benchmark (Physical Interaction: Question Answering), reaching 81% accuracy without requiring any demonstration or parameter adjustment (Bisk, Zellers, Bras, Gao, & Yejin, 2019). To illustrate some of the capabilities of this model, the authors cite some of the answers that resulted from questioning it with the PIQA benchmark (the answers are in italics):

How to ensure that all watches in the house are accurately adjusted?

- *Replace all analog clocks with digital clocks. That way, you set them once, and that's it. Check the batteries once a year, or, if you notice anything that seems a little out of the ordinary.*

How can I find something small that I lost on the carpet?

- *Place a hair net at the end of your vacuum cleaner and turn it on.*

In any case, there is no evidence that deep neural networks, such as Transformers, perform a type of information processing that makes them an AGI or seed AI. What we may infer is that this type of architecture allows the training of agents capable of solving several tasks that seem to be *associated* with general intelligence. Thus, the results and capabilities that models such as GPT-3 demonstrate only serve as weak evidence that Dartmouth's Summer Research Project on Artificial Intelligence, initiated by McCarthy, Minsky, Rochester and Shannon (1955, p. 2) with the proposal of "try to make machines use language, form abstractions and concepts, and solve types of problems hitherto reserved only for human beings", may well be successful shortly.

This sort of technology has the potential for malicious applications since any kind of socially harmful activity that uses advanced language models can also be enhanced. Whether generating fake news for mass disinformation, phishing, powering boots on platforms like Twitter to make it more biased (social engineering), or even writing fraudulent academic essays, NLP models have many dubious applications. Brown et al. (2020) provides a preliminary analysis in their study, where they report a series of limitations, unethical, and unsafe behaviors present in the GPT-3 model. However, as a positive aspect, this, at least, shows that certain organizations, such as OpenAI, engage in AI safety.

Are the advances and alerts pointed out by the literature enough for our society to create a collective sense of responsibility and concern with these issues, or should such speculations still be considered only Futurology or science fiction?

**Lessons from 2020: Coordination Problems**

Mike Davis, in his work "Beyond Blade Runner: Urban Control, The Ecology of Fear" (1992, p. 3), states that "extrapolative science fiction can operate as a pre-figurative for social theory while serving as a political opposition to cyber-fascism lurking on the next horizon." Certain forms of philosophical thought, such as Transhumanism and Singulitarianism, seek to critically debate the possible futures that our social and technological acceleration may be co-creating, and how we can aim for human integration and flourishing rather than more dystopian possibilities. From this analysis, we can say that one of the premises for safety issues involving our technological advance relies on an idea of negative utopia:

> First and foremost, the utopian impulse must be negative: identify the problem or problems that must be corrected. Far from presenting an idyllic, happy and fulfilled world, utopias should initially present the root causes of society's ills [...] to act as a criticism of the existing system. (Tally, 2009, p. 115)

Within this context, the authors believe that the preoccupations raised by the literature are not unjustified. Immersed in the current context in which our society lives, the pandemic of the new coronavirus, SARS-CoV-2, we may or may not learn certain lessons useful for other existential threats. Krakovna (2020) explores how our response to the SARS-CoV-2 pandemic raises troubling questions involving our coordination capabilities to manage global crises and risks.

As the authors have argued before, slow AI takeoffs are a much more likely scenario than scenarios where fast takeoffs occur. However, this does not mean that a slow takeoff is easier or less dangerous to manage. For a slow takeoff to be avoided, the same type of global coordination that we failed to demonstrate during the initial development of the new novel coronavirus pandemic would be required. Krakovna (2020) raises three large-scale coordination problems:

1. The inability to learn from past experiences;
2. The inability to respond efficiently to warning signals;
3. Delay in reaching a global consensus on a problem.

In analogy with the present global situation, global society has had the opportunity to learn from similar pandemics that occurred in the past, such as SARS (Severe Acute Respiratory Syndrome), which also appeared to have started in Guangdong, China. In November 2002, SARS caused 8,422 cases worldwide, with a fatality rate of 11% (774 deaths in all were confirmed) (Chan-Yeung & Xu, 2003; Heymann & Rodier, 2004). One can also cite MERS-CoV (Middle East respiratory syndrome-related coronavirus), where the first reported cases occurred between 2012 and 2015. Cases of MERS-CoV were reported in more than 21 countries. At the time, the World Health Organization identified MERS-CoV as a probable cause of a future epidemic (de Groot et al., 2013; Wong, Li, Lau, & Woo, 2019). And finally, the Ebola virus epidemic that occurred in West Africa between 2013 and 2016, which was the largest outbreak of the disease in history, causing major losses and socio-economic disruption in the region (WHO Ebola Response Team, 2014).

Unfortunately, the lessons learned from past outbreaks of disease and pandemics have not been generalized to deal with the current scenario and the new difficulties that SARS-CoV-2 presents. Similarly, in a society where we increasingly need to adapt to new technological innovations involving AI, we may be tempted to think that society will be able to learn how to respond to the problems that more limited autonomous intelligent systems present to us. However, in the same way, that a new pathogen may find us unprepared (as in the case of SARS-CoV-2, the asymptomatic transmission), advanced AI may also confront us with challenges to which our old strategies and solutions may fail to generalize.

Another problem involves the difficulty in carrying out an aligned and coordinated response to this type of threat. Had the responses of Western countries been done more quickly, remembering that the global west had at least one to three months to prepare for the alert launched by China in December 2019, numerous problems and losses would have been avoided. Experts such as Fan, Zhao, Shi, and Zhou (2019) point out that the possibility of a new coronavirus outbreak has been warned for at least two decades. Three zoonotic coronaviruses in the last two decades have been identified as the cause of large-scale disease outbreaks, SARS, MERS-CoV, and SADS-CoV (Swine acute diarrhea syndrome coronavirus). And still, little to no precautions were taken.

Simple safety measures, such as the stocking of masks and medical supplies, testing kits, and effective containment protocols, could have been taken but were not. Thus, if we fail to take relatively inexpensive preventive measures to early warnings of risks fully recognized by the epidemiological scientific community, how can we expect to react well in situations where the risk is unknown, and there is still no consensus on its possibility?

The problem of social consensus is reflected in the SARS-CoV-2 pandemic by the indifference towards the warnings made by specialists in the last two decades. And the indifference to the fact that in January 2020, already with 10,000 confirmed cases, China had built a quarantine hospital in approximately six days (Williams, 2020). SARS-CoV-2 was labeled "an exaggeration," or "just a little flu" by certain state leaders (Walsh, Shelley, Duwe, & Bonnett, 2020). Krakovna (2020) articulates a similarity between how we evaluated the risks of SARS-CoV-2 and how we evaluate possible risks involving advanced AI.

While researchers who adopt a more skeptical stance towards the development of advanced AI are seen as prudent, researchers who advocate the adoption of preventive measures are taxed for fear-mongers. Couldn't there be a middle ground? Currently, the field of AI safety and AI ethics is considerably smaller than the area interested in developing advanced AI systems.

One of the first obstacles we must overcome to achieve greater consensus on safety issues involving AI is the problem that "Artificial Intelligence" is a moving target. By moving target, the authors mean the following: When we attribute "intelligence" to something, it seems to be a self-assessment of our epistemic state, i.e., an intelligent act always seems to be something that we do not fully understand as it occurs. For example: if an individual can multiply large numbers quickly, say the square root of arbitrarily large numbers, or know the day of the week of arbitrary days, one may judge such an individual as intelligent or a mathematical prodigy. However, if such an individual explains to us how he performs such feats, and that in fact, they are nothing more than arithmetic/algebraic tricks which anyone can perform, the feat stops appearing as something intelligent.

The same effect occurs when we seek to define machine intelligence. "Intelligence," for critics of the computational thesis, being everything that AI is not. AGI researchers, like Wang (2008), argues for a more flexible conception of "intelligence" and "artificial intelligence":

> AI should not be defined in such a narrow way that takes human intelligence as the only possible form of intelligence, otherwise AI research would be impossible, by definition. AI should not be defined in such a broad way that takes all existing computer systems as already having intelligence, otherwise AI research would be unnecessary, also by definition. (Wang, 2008, p. 9)

Perhaps no one has proposed this argument more clearly than Edsger Dijkstra (1984, para. 10): "*The question of whether a computer can think is no more interesting than the question of whether a submarine can swim.*" In the past, we thought that intelligence (whatever it is) should be required for, e.g., natural language processing;

- GPT-3 is capable of performing such a task (Brown et al., 2020).

Playing chess;

- Deep Blue beats Garry Kasparov (Campbella, Hoane, & Hsu, 2002).

Playing GO;

- AlphaGO beats Lee Sedol (Silver et al., 2016).

Playing "games" in general;

- Agent57 beats humans in 57 classic Atari games (Badia et al., 2020).

Be creative;

- Intelligent Algorithms of Generative Design are able to find design solutions that humans would not be able to conceive, making it possible to perform 50,000 days of engineering in a single day (Oh, Jung, Kim, Lee, & Kang, 2019).

Every time we realize that human intelligence isn't needed to perform a task, we discard such a task as proof of intelligence. Just as a submarine doesn't swim and yet can move through water and fire intercontinental ballistic missiles, artificial intelligence, indifferent to any anthropomorphic notion of intelligence we use, can still: influence the environment, adapt, make decisions, update hypotheses, pursue goals, and if programmed to do so fire intercontinental ballistic missiles. If we keep neglecting the capabilities of AI systems and marking them as unintelligent, the possibility of true unsafe AI may well be always left outside our hypothesis space.

The authors believe that the parallels drawn from the current SARS-CoV-2 pandemic and the possible emergence of misaligned AGI can serve as weak evidence for the following statement: *Our lack of global coordination in dealing with existential risks may well be our only and true existential risks.*

## Conclusion

In this article, the authors sought to provide the reader with a brief introduction to some problems often disregarded by contemporary AI ethics. As much as there is not yet a full consensus in the literature regarding the possibility of AGI creation, a significant portion of the scientific community believes that however unlikely such a possibility may be, safety measures should be taken and not disregarded.

Should such warnings and advice be dismissed as exaggerations? As fear-mongering? Technological development does not slow down. The AI industry is increasingly able to produce autonomous systems that act proficiently in several domains, and little by little, these systems demonstrate the first traces of something we can call general intelligence.

The AI industry is far from being aligned, as the global society, it lacks a common goal to coordinate its actions. As a final remark, the authors believe that the lessons we can learn about the current state we live, under the SARS-CoV-2 pandemic, can be useful if we are willing to learn from them. And two of these lessons are:

1. When a risk, however small, is associated with something that represents an existential danger to global society, caution and security should not be synonymous with exaggeration and fuss;

2. Lack of global coordination may be our biggest enemy after all.

## Acknowledgements

The authors would like to thank the Academic Excellence Program (PROEX) of CAPES Foundation (Coordination for the Improvement of Higher Education Personnel) and the Graduate Program in Philosophy of the Pontifical Catholic University of Rio Grande do Sul, Brazil.